\documentclass[a4paper,11pt]{article}
\usepackage{amsmath,amssymb,amsthm,mathtools}
\usepackage{booktabs}
\usepackage{graphicx}
\usepackage[dvipsnames]{xcolor}

\makeatletter
\IfFileExists{jheppub.sty}{\usepackage{jheppub}}{%
  \usepackage[margin=2.5cm]{geometry}
  \usepackage[colorlinks=true,linkcolor=blue!60!black,citecolor=blue!60!black,urlcolor=blue!60!black]{hyperref}
  \renewcommand{\author}[2][]{\gdef\@author{##2\textsuperscript{##1}}}
  \def\@affils{}
  \newcommand{\affiliation}[2][]{\g@addto@macro\@affils{\par\textsuperscript{##1}##2}}
  \def\@email{}
  \newcommand{\emailAdd}[1]{\gdef\@email{\par\emph{E-mail:} \texttt{##1}}}
  \def\@abstr{}
  \renewcommand{\abstract}[1]{\gdef\@abstr{##1}}
  \def\@kw{}
  \newcommand{\keywords}[1]{\gdef\@kw{##1}}
  
  \renewcommand{\maketitle}{%
    \begin{center}
      {\LARGE\bfseries\@title\par}\vspace{1.2em}
      {\large\@author\par}\vspace{0.6em}
      {\small\itshape\@affils\par}\vspace{0.3em}
      {\small\@email\par}
    \end{center}\vspace{1em}
    \noindent\textsc{Abstract:} \@abstr\par\vspace{0.6em}
    \noindent\textsc{Keywords:} \@kw\par\vspace{1em}
    \hrule\vspace{0.5em}\tableofcontents\vspace{0.5em}\hrule\vspace{1em}}
}
\makeatother

\newtheorem{proposition}{Proposition}[section]
\newtheorem{lemma}[proposition]{Lemma}
\newtheorem{theorem}[proposition]{Theorem}
\newtheorem{corollary}[proposition]{Corollary}
\theoremstyle{remark}
\newtheorem{remark}[proposition]{Remark}

\newcommand{\ld}{\dot\ell}
\newcommand{\ldd}{\ddot\ell}
\newcommand{\tm}{\tau_{\max}}
\newcommand{\dd}{\mathrm{d}}
\newcommand{\Lg}{\mathcal L}
\newcommand{\Ham}{\mathcal H}
\newcommand{\Uc}{\mathcal U}
\newcommand{\Ic}{\mathcal I}
\newcommand{\Jc}{\mathcal J}
\newcommand{\Dc}{\mathcal D}
\newcommand{\Fc}{\mathcal F}
\newcommand{\Qt}{\tilde Q}
\newcommand{\Eh}{\hat E}

\title{Kerr--Schild Double Copy and Black-Hole Interiors from the Courant–Hilbert Approach}

\author[a]{H.~Babaei-Aghbolagh}
\affiliation[a]{Institute of Fundamental Physics and Quantum Technology, \& School of Physical Science and Technology, Ningbo University, Ningbo, Zhejiang 315211, China}
\emailAdd{hosseinbabaei@nbu.edu.cn}

\abstract{We study point charges of causal, electric--magnetic duality-invariant
nonlinear electrodynamics written in the Courant--Hilbert (CH)
representation, with an arbitrary CH-function $\ell(\tau)$. We consider
them from two gravitational viewpoints: the classical Kerr--Schild
double copy and Einstein gravity with a cosmological constant. The
near-origin behaviour of the electrostatic potential,
$A_t=c_0+c_1r+\dots$, is fixed by two theory data: a single functional,
$c_0=\sqrt Q\,\Ic[\ell]$, and the end point $\tm$ of the domain of
$\ell$, which gives $c_1=-\sqrt{2\tm}$. For the double copy, the Ricci
tensor is proportional to $\ddot\ell$. We prove a no-go theorem: no
causal theory yields a regular double-copy metric, and Born--Infeld-type
theories give the mildest singularity. For Einstein--NLED black holes,
the metric function is an algebraic function of $A_t$ and the energy
density for \emph{any} theory. The point-charge self-energy is
$\Uc_0=\frac43Qc_0$ and the next coefficient is $\Uc_1=2Qc_1$. The same
two coefficients therefore control the double copy and the horizon
structure: the Schwarzschild-like, Reissner--Nordstr\"om-like and
naked-singularity branches, and the star point. At small coupling, the
van der Waals critical point depends only on the first weak-field
coefficient of $\ell$. We give closed, often parametric, solutions for
ModMax, generalised Born--Infeld, logarithmic, $q$-deformed and
power-law (``no maximum-$\tau$'') electrodynamics. We express their
self-energies through $\Gamma(\tfrac14)$ where possible.

}


\begin{document}
\maketitle
\flushbottom

\section{Introduction}\label{sec:intro}

The double copy relates gauge theory to gravity. It originated in the KLT relations of string theory and in the colour--kinematics (BCJ) duality of scattering amplitudes, and has grown into a broad programme reviewed in~\cite{Bern:2019prr,Adamo:2022dcm}. A particularly transparent version is the \emph{classical} Kerr--Schild double copy of Monteiro, O'Connell and White~\cite{Monteiro:2014cda}. Stationary Kerr--Schild metrics $g_{\mu\nu}=\eta_{\mu\nu}+\phi\,k_\mu k_\nu$, with $k$ null and geodesic, are paired with abelian gauge fields $A_\mu=\phi\,k_\mu$. The Coulomb field of a point charge then maps to the Schwarzschild metric. When the gauge-theory side is allowed its full double-copy content, which includes a dilaton, the image of the point charge is the Janis--Newman--Winicour family~\cite{Kim:2019jwm}.

Duality-invariant nonlinear electrodynamics (NLED) has been formulated
through several complementary auxiliary-field approaches. An early and
particularly influential construction was introduced by Ivanov and Zupnik
(IZ), in which the nonlinear self-duality condition is explicitly encoded
in an interaction function of auxiliary fields, thereby providing a
systematic representation of a broad class of duality-invariant
theories~\cite{Ivanov:2002ab}. A manifestly Lorentz- and
duality-covariant extension, which unifies the auxiliary-tensor
formulation with the Pasti--Sorokin--Tonin (PST) approach~\cite{Pasti:1995tn}, was
subsequently developed by Ivanov, Nurmagambetov, and Zupnik
(INZ)~\cite{Ivanov:2014nya}. Independently, Avetisyan, Evnin, and Mkrtchyan introduced the
democratic, or clone-field, formulation, in which electric and magnetic
potentials enter on an equal footing~\cite{Avetisyan:2021heg,Avetisyan:2022zza}.
More recently, the relation between the INZ and clone-field formulations
was clarified through a systematic comparison of different descriptions
of nonlinear self-duality. Another useful perspective is provided by the
Courant--Hilbert (CH) description~\cite{Courant:Hilbert} of
duality-invariant NLED: the construction of Russo and
Townsend~\cite{Russo:2024ptw} can be related to the earlier approach of
Mkrtchyan and Svazas~\cite{Mkrtchyan:2022ulc}, which reconstructs
duality-invariant theories from their electrostatic solutions. Finally,
the foundational contributions of Bia\l ynicki-Birula to Born--Infeld
electrodynamics, its strong-field limit, and more general nonlinear
theories deserve particular recognition. Notably, the duality constraint
later associated with Gaillard--Zumino~\cite{Gaillard:1981rj,Gaillard:1997zr} and Gibbons--Rasheed~\cite{Gibbons:1995ap,Gibbons:1995cv} was already
obtained in an explicit form in Bia\l ynicki-Birula's 1983
work~\cite{bialynicki1983nonlinear}. Related developments can also be
found in Refs.~\cite{Gibbons:1995cv}.

All these constructions start from \emph{linear} Maxwell fields. Yet in string theory the effective action of an abelian open-string vector field is the Born--Infeld (BI) action~\cite{Born:1934gh,Fradkin:1985qd}, up to derivative corrections. The BI point charge has a finite electric field at the origin, a well-known example of how $\alpha'$-effects smooth out point-like singularities~\cite{Tseytlin:1995uq}. Pasarin and Tseytlin~\cite{Pasarin:2020qoa} therefore asked whether the double copy of the BI point charge gives a regular generalisation of the Schwarzschild metric. The metric components are indeed regular, but the curvature invariants still diverge at $r=0$. The culprit is the non-analytic term $\propto r=\sqrt{x_ix_i}$ in the potential, which is the constant BI field at the origin.

Born--Infeld theory is distinguished by much more than its string origin. It is invariant under electric--magnetic duality rotations~\cite{Gaillard:1997rt,Gibbons:1995cv,Tseytlin:1996it}. Duality-invariant nonlinear electrodynamics (NED) forms a large class, specified by a function of one variable~\cite{Gibbons:1995cv}. Its unique conformal member is ModMax~\cite{Bandos:2020jsw}. In the Courant--Hilbert (CH) representation~\cite{Russo:2024llm,Russo:2024xnh}, every such theory is generated by a single CH-function $\ell(\tau)$. Causality reduces to the simple conditions $\ld\ge1$ and $\ldd\ge0$~\cite{Russo:2024llm}. The same representation organises the $T\bar T$- and root-$T\bar T$-deformations of Maxwell and ModMax theory~\cite{Babaei-Aghbolagh:2022uij,Babaei-Aghbolagh:2025cni}. Born--Infeld, generalised BI (GBI), logarithmic, $q$-deformed and ``no maximum-$\tau$'' electrodynamics all appear as particular CH-functions~\cite{Russo:2024llm,Russo:2024xnh,Babaei-Aghbolagh:2025tim}.

Nonlinear electrodynamics is also a classic source for black holes with modified interiors. It underlies the regular black holes of Ay\'on-Beato and Garc\'ia~\cite{Ayon-Beato:1998hmi} and the families constructed in~\cite{Balart:2014cga}. More recently, Hale, Hennigar and Kubiz\v{n}\'ak~\cite{Hale:2025ezt} emphasised the role of the point-charge self-energy. When it is finite, the Cauchy horizon of Reissner--Nordstr\"om is no longer compulsory, and the interior is Schwarzschild-like whenever the mass exceeds the self-energy. Russo and Townsend~\cite{Russo:2026vnj} classified the near-origin behaviour of self-dual NED and its black holes. In~\cite{Babaei-Aghbolagh:2025tim}, AdS black holes of the root-$T\bar T$-generated GNED family were analysed, with emphasis on thermodynamics and on naked-singularity criteria. See more in~\cite{Fathi:2026ctx,Ovgun:2025mdg,Mendez-Zavaleta:2026rgg,Sucu:2026nkw,Guo:2026xlk,Guzman:2026muu,Zhou:2026ako}. The finiteness of the point-charge self-energy in BI-type theories has also been examined from a field-theory perspective~\cite{Dehghani:2021fwb}.

In this paper we bring these two lines together. We ask what the CH-function $\ell(\tau)$ determines about
\begin{enumerate}
\item the Kerr--Schild double copy of the point charge, and
\item the interior and thermodynamics of the Einstein--NED black hole sourced by the same charge.
\end{enumerate}
Our main results are:
\begin{itemize}
\item \textbf{Two numbers control the origin.} The potential behaves as $A_t=c_0+c_1r+\dots$ with $c_0=\sqrt Q\,\Ic[\ell]$, where $\Ic[\ell]$ is an explicit integral~\eqref{Ic}. For every theory with a finite $\tm$, $c_1=-\sqrt{2\tm}$ is independent of the charge.
\item \textbf{Double copy.} The Ricci tensor of the double-copy metric is proportional to $\ldd$. It vanishes only at the conformal (ModMax) point. A no-go theorem shows that no causal theory gives a regular double copy. In a natural gauge, BI-type theories leave the universal residual singularity $K\simeq16\tm/r^2$ with a regular Weyl tensor.
\item \textbf{Einstein--NED black holes.} For \emph{any} NED, $f(r)=k-m/r-\Lambda r^2/3+\tfrac43QA_t/r-\tfrac23r^2\Ham$, so the Einstein equations are solved algebraically once $A_t$ is known. The near-origin data are $\Uc_0=\tfrac43Qc_0$ and $\Uc_1=2Qc_1$. These decide between the Schwarzschild-like, Reissner--Nordstr\"om-like and naked branches, and fix the star point $2Q\sqrt{2\tm}=k$.
\item \textbf{Thermodynamics.} The first law holds with $\Phi\propto A_t(r_h)$ for every theory. At $O(\lambda)$ the critical point depends only on the first weak-field coefficient $h_1$ of $\ell$.
\item \textbf{Examples.} We give closed or parametric solutions for ModMax, GBI, logarithmic, $q$-deformed ($\alpha$-family) and power-law (``no maximum-$\tau$'') theories. Their self-energies are expressed in $\Gamma(\tfrac14)$ where possible.
\end{itemize}

The paper is organised as follows. Section~\ref{sec:CH} reviews the CH representation and derives three lemmas on point charges that are used throughout. Section~\ref{sec:DC} studies the Kerr--Schild double copy. Section~\ref{sec:BH} studies Einstein--NED black holes in (A)dS and their thermodynamics. Section~\ref{sec:ex} treats the examples. Section~\ref{sec:disc} contains the discussion. Appendices collect special-function identities (\ref{app:special}).

\paragraph{Notation.} We use $(-,+,+,+)$ signature and Gauss's law $r^2D=Q$, with $Q>0$. Symbols used throughout:
\begin{itemize}
\item $\Lg$ is the NED Lagrangian, $\Ham=ED-\Lg$ the electrostatic energy density, and $\Uc(r)$ the self-energy outside radius $r$.
\item $f(r)$ is the black-hole metric function and $\phi(r)$ the Kerr--Schild potential.
\item $k\in\{1,0,-1\}$ is the horizon curvature and $L$ the AdS radius.
\item $T_H$ and $S_{\rm BH}$ are the Hawking temperature and the entropy.
\item The ModMax and $T\bar T$ couplings are $\gamma\ge0$ and $\lambda>0$. We often use the effective charge $\Qt=e^{-\gamma/2}Q$ and the rescaled field $\Eh=e^{\gamma/2}E$.
\item $B(a,b)$ is the Euler Beta function (appendix~\ref{app:special}).
\end{itemize}

\section{Self-dual NED and its point charges}\label{sec:CH}

\subsection{Courant--Hilbert representation}

With $S=-\tfrac14F_{\mu\nu}F^{\mu\nu}=\tfrac12(\mathbf E^2-\mathbf B^2)$ and $P=-\tfrac14F_{\mu\nu}\tilde F^{\mu\nu}=\mathbf E\cdot\mathbf B$, introduce
\begin{equation}
U=\tfrac12\big(\sqrt{S^2+P^2}-S\big),\qquad V=\tfrac12\big(\sqrt{S^2+P^2}+S\big).
\end{equation}
Electric--magnetic duality invariance~\cite{Gaillard:1997rt,Gibbons:1995cv} becomes $\Lg_U\Lg_V=-1$. Its general solution is~\cite{Russo:2024llm}
\begin{equation}\label{CH}
\Lg(U,V)=\ell(\tau)-\frac{2U}{\ld(\tau)},\qquad \tau=V+\frac{U}{\ld(\tau)^2}.
\end{equation}
We normalise the CH-function as in~\cite{Babaei-Aghbolagh:2025tim},
\begin{equation}\label{weak}
\ell(\tau)=e^\gamma\tau+\sum_{i\ge1}h_i\,\lambda^i e^{(i+1)\gamma}\tau^{i+1},
\end{equation}
so that every theory reduces to ModMax~\cite{Bandos:2020jsw} as $\lambda\to0$ (and to Maxwell if also $\gamma=0$).\footnote{The coefficients $h_i$ are denoted $m_i$ in~\cite{Babaei-Aghbolagh:2025tim}. We reserve $m$ for the black-hole mass parameter.}

Causality and convexity require~\cite{Russo:2024llm}
\begin{equation}\label{causal}
\ld\ge1,\qquad \ldd\ge0 .
\end{equation}
The trace of the stress tensor is $\Theta=T^\mu{}_\mu=4(\ell-\tau\ld)$~\cite{Russo:2024xnh,Babaei-Aghbolagh:2025tim}, hence
\begin{equation}\label{trace}
\frac{\dd\Theta}{\dd\tau}=-4\tau\,\ldd .
\end{equation}
Born--Infeld with tension $T_{\rm BI}$ corresponds to $\gamma=0$ and $\lambda=1/T_{\rm BI}$, and generally we identify $T\equiv1/\lambda$. The tension of~\cite{Pasarin:2020qoa} is $T_{\rm PT}=T^{1/2}$.

\subsection{Electrostatic point charges}

For a static, purely electric, spherically symmetric field, $P=0$, $U=0$ and $\tau=V=S=E^2/2$. Hence the CH-function \emph{is} the Lagrangian on the electric slice:
\begin{equation}\label{Dc}
\Lg\big|_{\mathbf B=0}=\ell(\tfrac12E^2),\qquad
D=\Dc(E)\equiv E\,\ld(\tfrac12E^2),\qquad
\Dc'(E)=\ld+2\tau\ldd\ \ge1 .
\end{equation}
Gauss's law $\Dc(E)=Q/r^2$ therefore has a unique solution, with
\begin{equation}\label{Eprime}
E'(r)=-\frac{2Q}{r^3(\ld+2\tau\ldd)}<0 .
\end{equation}
So $E$ is positive and strictly decreasing, in agreement with the general analysis of~\cite{Russo:2026vnj}. At large $r$, $E\simeq e^{-\gamma}Q/r^2$.

Let $\tm\in(0,\infty]$ be the end of the domain of $\ell$. A point-charge solution valid for all $r>0$ exists if and only if $\Dc$ is unbounded. This leaves two classes:
\begin{itemize}
\item[\textbf{A}] (Born--Infeld class) $\tm<\infty$ with $\ld\simeq\mathcal A\,(\tm-\tau)^{-\kappa}$, $\kappa>0$. The field at the origin is finite, $E(0)=E_0\equiv\sqrt{2\tm}$.
\item[\textbf{B}] $\tm=\infty$ with $\ell\simeq\ell_\infty\tau^\beta$ as $\tau\to\infty$; causality requires $\beta\ge1$. The field diverges at the origin.
\end{itemize}
A third possibility, $\tm<\infty$ with $\ld(\tm)$ finite, needs an extended source~\cite{Mkrtchyan:2022ulc} and is not considered.

\subsection{Three lemmas}

Define
\begin{equation}\label{Ic}
\Jc(\tau)\equiv\int_0^{\tau}\frac{\dd\tau'}{(2\tau')^{3/4}\,\ld(\tau')^{1/2}},\qquad \Ic[\ell]\equiv\Jc(\tm).
\end{equation}

\begin{lemma}[Potential]\label{lem:At}
With $A_t(\infty)=0$,
\begin{equation}\label{Atgen}
A_t(r)=\sqrt Q\,\Jc\big(\tau(r)\big)-r\,E(r).
\end{equation}
Whenever $rE\to0$ at the origin, the value at the origin is
\begin{equation}\label{c0}
c_0\equiv A_t(0)=\sqrt Q\,\Ic[\ell].
\end{equation}
This holds always in case~A, and in case~B iff $\beta>\tfrac32$.
\end{lemma}
\begin{proof}
$A_t=\int_r^\infty E=-rE+\int_0^{E(r)}r(E')\,\dd E'$ with $r(E)=\sqrt{Q/\Dc(E)}$. Substituting $E=\sqrt{2\tau}$ gives~\eqref{Atgen}. In case~B the integrand of~\eqref{Ic} behaves as $\tau^{-3/4-(\beta-1)/2}$ at large $\tau$.
\end{proof}

\begin{lemma}[Near-origin field]\label{lem:origin}
In case~A,
\begin{equation}\label{expA}
E=E_0-e_1\,r^{\,n-1}+\dots,\qquad
A_t=c_0+c_1r+\frac{e_1}{n}\,r^{\,n}+\dots,
\end{equation}
with
\begin{equation}
c_1=-E_0=-\sqrt{2\tm},\qquad n=1+\frac2\kappa,\qquad e_1=\frac{1}{E_0}\Big(\frac{\mathcal A E_0}{Q}\Big)^{1/\kappa}.
\end{equation}
In case~B,
\begin{equation}
E\simeq\mathcal C\,r^{-p},\qquad p=\frac{2}{2\beta-1}\in(0,2],\qquad \mathcal C=\Big(\frac{2^{\beta-1}Q}{\ell_\infty\beta}\Big)^{\frac{1}{2\beta-1}} .
\end{equation}
\end{lemma}
\begin{proof}
In case~A, $\mathcal AE_0(\tm-\tau)^{-\kappa}\simeq Q/r^2$ and $E=\sqrt{2\tau}\simeq E_0-(\tm-\tau)/E_0$. Case~B is immediate.
\end{proof}
Thus $c_1$ is independent of $Q$, $\mathcal A$ and $\kappa$, and cannot vanish in case~A. The exponent $n$ is in general not an integer; Born--Infeld ($\kappa=\tfrac12$) has $n=5$.

\begin{lemma}[Virial identity]\label{lem:virial}
For any NED point charge,
\begin{equation}\label{virial}
2\int_r^\infty s^2\Ham(s)\,\dd s=\frac43\,Q\,A_t(r)-\frac23\,r^3\Ham(r).
\end{equation}
\end{lemma}
\begin{proof}
$\Ham$ is the Hamiltonian density $H(D)$, so $\Ham'=E\,D'=-2QE/r^3$. Differentiating the right-hand side then gives $-2r^2\Ham$, and both sides vanish at infinity.
\end{proof}

\section{Kerr--Schild double copy}\label{sec:DC}

\subsection{Metric and curvature}

Following~\cite{Monteiro:2014cda,Pasarin:2020qoa}, gauge-transform the point charge to $A_\mu=\phi(r)k_\mu$ with $k_\mu=(1,x_i/r)$ null, where $\phi=A_t$. The Kerr--Schild metric $g_{\mu\nu}=\eta_{\mu\nu}+\phi\,k_\mu k_\nu$ becomes, in Schwarzschild coordinates,
\begin{equation}\label{DCmetric}
\dd s^2=-(1-\phi)\,\dd t^2+\frac{\dd r^2}{1-\phi}+r^2\dd\Omega^2 .
\end{equation}
At large $r$ this is Schwarzschild with $M_{\rm DC}=\tfrac12e^{-\gamma}Q$. For general $\phi$, with $\phi'=-E$,
\begin{align}
R&=\frac{(r^2\phi)''}{r^2}=\frac{2\phi}{r^2}-\frac{4E}{r}-E',\label{R}\\
K\equiv R_{\mu\nu\rho\sigma}R^{\mu\nu\rho\sigma}&=\frac{4\phi^2}{r^4}+\frac{4E^2}{r^2}+E'^2,\label{K}\\
C_{\mu\nu\rho\sigma}C^{\mu\nu\rho\sigma}&=\frac{(2\phi+2rE-r^2E')^2}{3r^4}.\label{C}
\end{align}

\subsection{Near-origin structure}

Near the origin $\phi=c_0+c_1r+\dots$, with $c_{0,1}$ given by Lemmas~\ref{lem:At} and~\ref{lem:origin}. In case~A, inserting~\eqref{expA} gives
\begin{align}
R&=\frac{2c_0}{r^2}-\frac{6E_0}{r}+e_1\Big(n+3+\frac2n\Big)r^{\,n-2}+\dots,\label{RA}\\
K&=\frac{4c_0^2}{r^4}-\frac{8c_0E_0}{r^3}+\frac{8E_0^2}{r^2}+\frac{8c_0e_1}{n}\,r^{\,n-4}+\dots,\label{KA}\\
C^2&=\frac{4c_0^2}{3r^4}+\frac{4c_0}{3}\frac{(n-1)(n-2)e_1}{n}\,r^{\,n-4}+\dots,\label{CA}
\end{align}
where the $r^{-3}$ term of $C^2$ cancels identically. The constant $c_0$ produces a solid-angle defect and can be removed by the constant gauge shift $\phi\to\tilde\phi=\phi-c_0$, at the price of non-flat asymptotics. The term $c_1r=c_1\sqrt{x_ix_i}$ cannot be removed~\cite{Pasarin:2020qoa}. In the shifted gauge
\begin{equation}\label{shift}
\tilde K=\frac{8E_0^2}{r^2}+\dots=\frac{16\,\tm}{r^2}+\dots,\qquad
\tilde R=-\frac{6E_0}{r}+\dots,\qquad
\tilde C^2=\frac13\Big[\frac{(n-1)(n-2)e_1}{n}\Big]^2r^{2n-4}+\dots
\end{equation}
The residual singularity thus has a universal coefficient fixed by $\tm$ alone. For $\kappa\le2$ (including all examples below) it is purely of Ricci type.

In case~B with $\beta>\tfrac32$ ($p<1$), $c_0$ is finite but $\phi=c_0-\frac{\mathcal C}{1-p}r^{1-p}+\dots$ is not even $C^1$. In every subcase, the non-constant part of $\phi$ gives
\begin{equation}\label{KB}
K_{\rm sing}=\mathcal C^2\Big[\frac{4}{(1-p)^2}+4+p^2\Big]r^{-2-2p},
\end{equation}
which reduces to $12Q^2/r^6$ for Maxwell ($p=2$). Case~A is the $p\to0$ limit.

\subsection{Ricci tensor, effective source and a no-go theorem}

Let $\varpi(r)\equiv r^2E=Q/\ld(\tau(r))$. By~\eqref{Eprime} and~\eqref{trace},
\begin{equation}\label{varpi}
\varpi'=\frac{2Q^3\,\ldd}{r^5\,\ld^{\,3}(\ld+2\tau\ldd)}=-\frac{Q^3}{2\tau\,r^5\,\ld^{\,3}(\ld+2\tau\ldd)}\frac{\dd\Theta}{\dd\tau}\;\ge0 .
\end{equation}

\begin{proposition}\label{prop:Ricci}
The Ricci tensor of~\eqref{DCmetric} is
\begin{equation}
R^t{}_t=R^r{}_r=-\frac{\varpi'}{2r^2},\qquad
R^\theta{}_\theta=R^\varphi{}_\varphi=\frac{\phi-rE}{r^2}=\frac1{r^2}\int_r^\infty\frac{\varpi'(s)}{s}\,\dd s .
\end{equation}
Consequently:
\begin{enumerate}
\item The double copy is Ricci-flat if and only if $\ldd\equiv0$, i.e.\ for ModMax/Maxwell. Its Ricci tensor measures $\dd\Theta/\dd\tau$, and vanishes exactly at the conformal fixed point of the root-$T\bar T$ flow.
\item The effective source is an anisotropic fluid with $\rho_{\rm eff}=-p_r\propto(\phi-rE)/r^2\ge0$ and $p_\theta\propto\varpi'\ge0$. Positivity is guaranteed by convexity, $\ldd\ge0$.
\end{enumerate}
\end{proposition}

\begin{theorem}[No-go]\label{thm:nogo}
Let $E(r)$ be the point-charge field of any causal NED, self-dual or not, so that $E>0$ and $E$ is non-increasing. Then the metric~\eqref{DCmetric} obeys
\begin{equation}
K\ge\max\Big(\frac{4\phi^2}{r^4},\frac{4E^2}{r^2}\Big),
\end{equation}
so $K\to\infty$ at the origin for \emph{every} additive constant in $\phi$. With flat asymptotics, $K\ge4c_0^2r^{-4}[1+o(1)]$. In the shifted gauge, $K\ge4E(r_1)^2/r^2$ for all $r<r_1$.
\end{theorem}
The bounds are saturated in power by case~A. Within the Kerr--Schild double copy, BI-type theories are therefore optimal, and a regular double copy would require $E(0)=0$, which is incompatible with~\eqref{Eprime}.

The condition $E(0)=0$ was identified in~\cite{Pasarin:2020qoa,Mkrtchyan:2022ulc} as necessary and sufficient for a regular double copy. Mkrtchyan and Svazas~\cite{Mkrtchyan:2022ulc} constructed electrostatic configurations with $E(0)=0$ and used them to build regular double-copy black holes. These configurations are of two kinds: solitons, and solutions with sources of finite size. The two kinds relate to Theorem~\ref{thm:nogo} differently.

\begin{corollary}[Solitons are acausal]\label{cor:soliton}
Let $\Lg(S,P)$ be any NED, self-dual or not, and let $E(r)$ be a static, spherically symmetric electric solution that satisfies $r^2D=Q$ for all $r>0$, with $E(0)=0$ and $E>0$ for $r>0$. Then on some interval of field strengths
\begin{equation}\label{acausal}
\frac{\dd D}{\dd E}=\Lg_S+2S\,\Lg_{SS}<0 ,
\end{equation}
so the theory violates the causality condition $\Lg_S+2S\Lg_{SS}>0$ for purely electric fields~\cite{Russo:2026vnj}. For CH theories this condition is $\Dc'(E)=\ld+2\tau\ldd\ge1$, see~\eqref{Dc}. In particular, the solitonic solutions of~\cite{Mkrtchyan:2022ulc} that give regular double-copy black holes cannot arise in a causal NED.
\end{corollary}
\begin{proof}
For $\mathbf B=0$ we have $D=\Lg_SE$ and $S=E^2/2$, so $\dd D/\dd E=\Lg_S+E^2\Lg_{SS}=\Lg_S+2S\Lg_{SS}$. Gauss's law makes $D=Q/r^2$ strictly decreasing in $r$. Since $E$ vanishes at $r=0$ and is positive at larger $r$, there is an interval on which $\dd E/\dd r>0$. On this interval $\dd D/\dd E=(\dd D/\dd r)/(\dd E/\dd r)<0$.
\end{proof}

\begin{figure}[t]
\centering
\includegraphics[width=\textwidth]{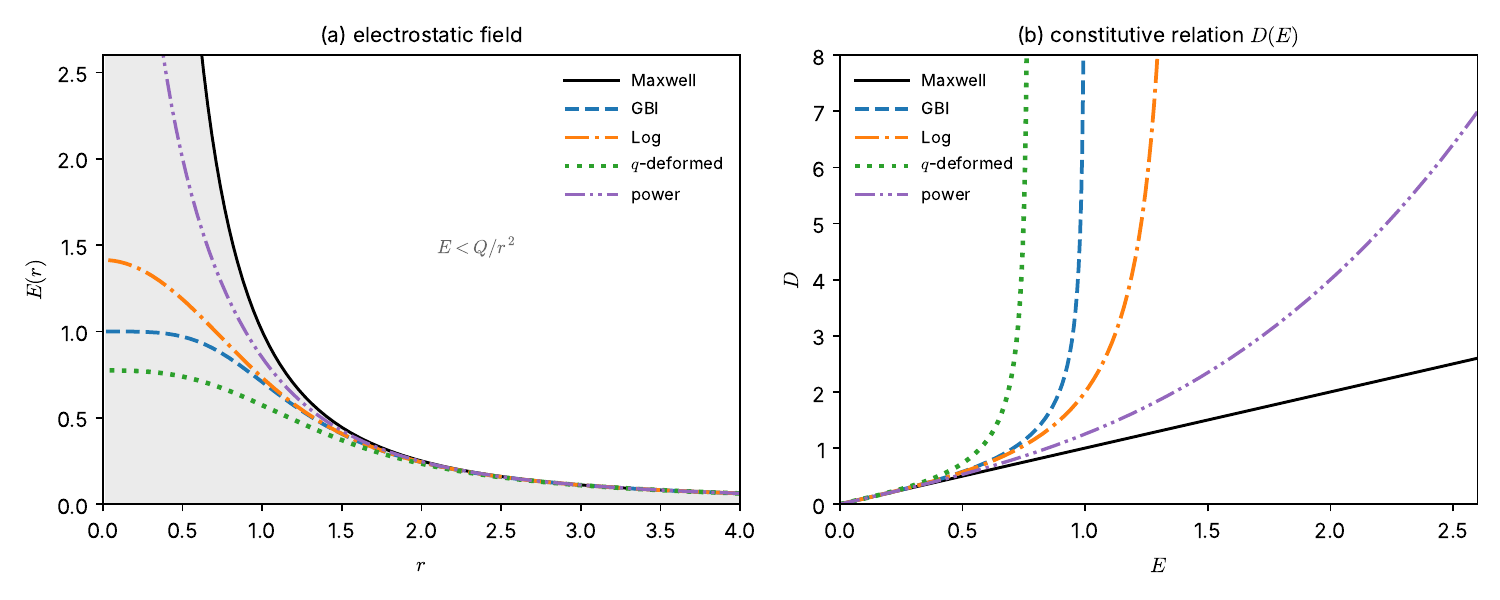}
\caption{Point charges of causal CH theories, for $Q=\lambda=1$ and $\gamma=0$.
(a)~The field $E(r)$. For every causal theory it is positive, strictly decreasing and below the Coulomb curve $Q/r^2$ (grey region)~\cite{Russo:2026vnj}. In case~A (GBI, Log, and $q$-deformed with $\alpha=0.3$) it reaches the finite value $E_0=\sqrt{2\tm}$ at the origin, so that $c_1=-E_0\neq0$ and Theorem~\ref{thm:nogo} applies. In case~B (power family with $\beta=2$) it diverges more slowly than Coulomb, as $r^{-p}$ with $p=\tfrac23$.
(b)~The constitutive relation $D(E)$. Causality requires $\dd D/\dd E=\Lg_S+2S\Lg_{SS}>0$, i.e.\ $\Dc'(E)=\ld+2\tau\ldd\ge1$, so every causal curve is single-valued and increasing. The case-A curves have vertical asymptotes at $E_0=\sqrt{2\tm}$. A configuration with $E(0)=0$ would require a branch with $\dd D/\dd E<0$ (Corollary~\ref{cor:soliton}).}\label{fig:dcfield}
\end{figure}

As an explicit example, consider the soliton (4.22) of~\cite{Mkrtchyan:2022ulc},
\begin{equation}
E(r)=\frac{Q\,r^2}{\sqrt{r^8+r_0^8}} .
\end{equation}
It rises for $r<r_0$, peaks at $r=r_0$ and decays as $Q/r^2$ for $r\gg r_0$. Hence $\dd D/\dd E<0$ for $r<r_0$, and $\dd D/\dd E$ diverges at $r=r_0$. The constitutive relation $D(E)$ is therefore double-valued, and no single-valued causal Lagrangian can support the whole solution. The same holds for the family $E\propto r^n\,(r^{(n+2)/m}+r_0^{(n+2)/m})^{-m}$ of~\cite{Mkrtchyan:2022ulc}, which has $E\propto r^n$ near the origin. By contrast, Figure~\ref{fig:dcfield} shows the point charges of causal CH theories, for which $E$ is monotonic and $D(E)$ is single-valued.

\begin{remark}[Finite-size sources]\label{rem:finite}
Configurations whose charge is spread over a region of finite size evade Theorem~\ref{thm:nogo} for a different reason. Inside the source, Gauss's law reads $r^2D=Q(r)$ with $Q(0)=0$, so $D$ can grow from zero and $E(0)=0$ is compatible with $\dd D/\dd E>0$ everywhere. What is lost is not causality but the point-like source: near the origin the single copy is no longer a vacuum NED solution, and the double copy is no longer the image of a point charge. This is the situation of case~C in section~\ref{sec:CH}. Within the Kerr--Schild double copy of a point charge, regularity and causality are therefore mutually exclusive.
\end{remark}


Since $1-\phi$ increases monotonically from $1-c_0$ to $1$, the metric~\eqref{DCmetric} has a single horizon if and only if $c_0>1$, i.e.
\begin{equation}\label{DChorizon}
Q>Q_*^{\rm DC}\equiv\Ic[\ell]^{-2}.
\end{equation}
Otherwise the singularity is naked. In case~B with $\beta\le\tfrac32$ there is always a horizon.

\section{Einstein--NED black holes}\label{sec:BH}

\subsection{Universal solution}

Consider~\cite{Babaei-Aghbolagh:2025tim}
\begin{equation}
I=\frac{1}{16\pi}\int\dd^4x\sqrt{-g}\,(R-2\Lambda)+\frac{1}{4\pi}\int\dd^4x\sqrt{-g}\,\Lg,\qquad
G_{\mu\nu}+\Lambda g_{\mu\nu}=2T_{\mu\nu},
\end{equation}
with $\Lambda=-3/L^2$ and
\begin{equation}
\dd s^2=-f\,\dd t^2+\frac{\dd r^2}{f}+r^2\dd\Omega_k^2 .
\end{equation}
For static electric fields $T^t{}_t=T^r{}_r=-\Ham$, and the $tt$ equation integrates to
\begin{equation}\label{fU}
f(r)=k-\frac mr-\frac{\Lambda r^2}{3}+\frac{\Uc(r)}{r},\qquad \Uc(r)=2\int_r^\infty s^2\Ham(s)\,\dd s .
\end{equation}
The mass is $M=2\Omega_km$, and $\Uc$ is the self-energy outside radius $r$ in the normalisation of~\cite{Hale:2025ezt}. Lemma~\ref{lem:virial} turns~\eqref{fU} into an algebraic statement.

\begin{proposition}[Exact solution]\label{prop:exact}
For any NED,
\begin{equation}\label{fexact}
f(r)=k-\frac mr-\frac{\Lambda r^2}{3}+\frac{4Q}{3}\,\frac{A_t(r)}{r}-\frac23\,r^2\Ham(r).
\end{equation}
\end{proposition}
Once $A_t$ is known, solving the Einstein equations is therefore never an obstruction. For CH theories $A_t$ is given by~\eqref{Atgen}.

\begin{proposition}[Near-origin data]\label{prop:U}
In case~A, and in case~B with $\beta>\tfrac32$,
\begin{equation}\label{U0}
\Uc_0\equiv\Uc(0)=\frac43\,Q\,c_0=\frac43\,Q^{3/2}\,\Ic[\ell].
\end{equation}
In case~A moreover
\begin{equation}\label{U1}
\Uc(r)=\Uc_0+\Uc_1r+o(r),\qquad \Uc_1=2Qc_1=-2Q\sqrt{2\tm},
\end{equation}
so that
\begin{equation}\label{rf}
\Fc(r)\equiv rf(r)=-\big(m-\Uc_0\big)+\big(k-2Q\sqrt{2\tm}\big)\,r+o(r).
\end{equation}
\end{proposition}
\begin{proof}
Set $r=0$ in~\eqref{virial}; in both cases $r^3\Ham\to0$. For~\eqref{U1}, use $r^2\Ham=QE-r^2\Lg\to QE_0$.
\end{proof}

All gravitational data depend on $(Q,\gamma)$ only through $\Qt$, because $QE=\Qt\Eh$ and $\Ham$ is a function of $\Eh$ alone. In particular $\Uc_1=-2\Qt\,\Eh_0$.

\begin{remark}[Bridge]
The two coefficients $(c_0,c_1)$ that control the double copy also fix $(\Uc_0,\Uc_1)$. Both constructions are governed by the single functional $\Ic[\ell]$ and the single number $\tm$.
\end{remark}

\subsection{Interior structure}

For $m\ne\Uc_0$, writing $\Delta\equiv m-\Uc_0$ and using $K=f''^2+4f'^2/r^2+4(f-k)^2/r^4$, case~A gives
\begin{equation}\label{KBH}
K=\frac{12\Delta^2}{r^6}+\frac{16\,QE_0\,\Delta}{r^5}+\frac{16\,Q^2E_0^2}{r^4}+\dots
\end{equation}
On the marginal surface $m=\Uc_0$ this softens to $K\simeq16Q^2E_0^2/r^4$.

The strong energy condition holds for causal self-dual NED ($\ell\ge0$~\cite{Russo:2026vnj}). It makes $\Fc$ convex~\cite{Hale:2025ezt}, so there are at most two horizons. With~\eqref{rf}, case~A then has three branches ($\Lambda\le0$):
\begin{itemize}
\item $m>\Uc_0$: exactly one horizon and a spacelike singularity (Schwarzschild-like branch, no Cauchy horizon).
\item $m<\Uc_0$ and $k>2Q\sqrt{2\tm}$: $\Fc(0)>0$ and $\Fc'(0)>0$, so the singularity is naked.
\item $m<\Uc_0$ and $k<2Q\sqrt{2\tm}$ ($k=1$): Reissner--Nordstr\"om-like branch, with event and Cauchy horizons for $m>m_e$ and a naked singularity for $m<m_e$.
\end{itemize}
The extremal mass follows from $\Fc=\Fc'=0$:
\begin{equation}\label{extremal}
2r_e^2\,\Ham(r_e)=k-\Lambda r_e^2,\qquad m_e=k\,r_e-\frac{\Lambda r_e^3}{3}+\Uc(r_e).
\end{equation}
The extremal curve meets $m=\Uc_0$ at $r_e\to0$. This defines the \emph{star point}
\begin{equation}\label{star}
2Q\sqrt{2\tm}=k\qquad(\Lambda=0).
\end{equation}
In case~B with $\beta>\tfrac32$, instead, $\Fc'(r)\to-\infty$ at the origin. There is no star point, and the RN-like branch exists for every charge. For $\beta\le\tfrac32$, $\Uc$ diverges, $f\to+\infty$ at the origin, and the Cauchy horizon is compulsory, as in Maxwell theory.

\subsection{Thermodynamics}

From $f(r_h)=0$,
\begin{equation}\label{TH}
m=k\,r_h-\frac{\Lambda r_h^3}{3}+\Uc(r_h),\qquad
T_H=\frac{f'(r_h)}{4\pi}=\frac{1}{4\pi}\Big(\frac{k}{r_h}-\Lambda r_h-2r_h\Ham(r_h)\Big).
\end{equation}
Since $\partial_Q\Ham|_r=E/r^2$, we have $\partial_Q\Uc(r_h)=2A_t(r_h)$. With $M=2\Omega_km$, $S_{\rm BH}=4\pi\Omega_kr_h^2$ and $\rho_Q=Q\Omega_k/4\pi$ as in~\cite{Babaei-Aghbolagh:2025tim}, this gives
\begin{equation}\label{firstlaw}
\dd M=T_H\,\dd S_{\rm BH}+\Phi\,\dd\rho_Q,\qquad \Phi=16\pi A_t(r_h),
\end{equation}
for every theory. The free energy is $F=M-T_HS_{\rm BH}=\Omega_k\big(2m-4\pi r_h^2T_H\big)$.

For $k=1$ and small $\lambda$, \eqref{TH} becomes
\[
4\pi L\,T_H=\frac1x+3x-\frac{y^2}{x^3}+\frac{\varepsilon\,y^4}{2x^7}+O(\lambda^2),\qquad
x=\frac{r_h}{L},\quad y=\frac{\Qt}{L},\quad \varepsilon=\frac{\lambda h_1}{L^2}.
\]
Solving $\partial_xT_H=\partial_x^2T_H=0$ perturbatively gives a universal critical point:
\begin{equation}\label{crit}
\Qt_c=\frac L6\Big(1+\frac{7\varepsilon}{12}\Big),\qquad
r_c=\frac{L}{\sqrt6}\Big(1-\frac{7\varepsilon}{12}\Big),\qquad
T_c=\frac{\sqrt6}{3\pi L}\Big(1-\frac{\varepsilon}{12}\Big)+O(\varepsilon^2).
\end{equation}
Theories with equal $h_1$ share the same critical point at $O(\lambda)$, irrespective of their strong-field behaviour.

\section{Examples}\label{sec:ex}

In all examples we use the parametrisation~\eqref{weak} with explicit $(\gamma,\lambda)$. Appendix~\ref{app:special} collects the Beta-function values.

\subsection{ModMax}
For $\ell=e^\gamma\tau$ we have $E=e^{-\gamma}Q/r^2$ and $\phi=e^{-\gamma}Q/r$. The double copy is exactly Schwarzschild with $M_{\rm DC}=\tfrac12e^{-\gamma}Q$, which is Ricci-flat (Proposition~\ref{prop:Ricci}). The Einstein solution is Reissner--Nordstr\"om with charge $\Qt$~\cite{Bandos:2020jsw,Babaei-Aghbolagh:2025tim}. The root-$T\bar T$ parameter $\gamma$ only renormalises the charge.

\subsection{Generalised Born--Infeld}
GBI~\cite{Babaei-Aghbolagh:2022uij,Babaei-Aghbolagh:2025tim} has
\begin{equation}
\ell=\frac1\lambda\Big[1-\sqrt{1-2e^\gamma\lambda\tau}\Big],\qquad \tm=\frac{e^{-\gamma}}{2\lambda},\qquad\kappa=\tfrac12,\qquad h_1=h_2=\tfrac12,\quad h_3=\tfrac58 .
\end{equation}
The field and potential are
\begin{equation}
E=\frac{e^{-\gamma}Q}{\sqrt{r^4+r_0^4}},\qquad r_0^4=\lambda e^{-\gamma}Q^2,\qquad
A_t=\frac{e^{-\gamma}Q}{r}\,{}_2F_1\Big(\tfrac14,\tfrac12;\tfrac54;-\tfrac{r_0^4}{r^4}\Big),
\end{equation}
and the near-origin data are
\begin{equation}
c_0=\frac{\Gamma(\frac14)^2}{4\sqrt\pi}\frac{e^{-\gamma/2}\Qt^{1/2}}{\lambda^{1/4}},\qquad
c_1=-\frac{e^{-\gamma/2}}{\sqrt\lambda},\qquad
\Uc_0=\frac{\Gamma(\frac14)^2}{3\sqrt\pi}\frac{\Qt^{3/2}}{\lambda^{1/4}},\qquad
\Uc_1=-\frac{2\Qt}{\sqrt\lambda} .
\end{equation}
These reproduce the near-origin expansion of~\cite{Babaei-Aghbolagh:2025tim}. For $\gamma=0$ and $\lambda=T_{\rm PT}^{-2}$ they also reproduce the Born--Infeld result of~\cite{Pasarin:2020qoa}.
\footnote{Eq.~(4.3) of~\cite{Pasarin:2020qoa} contains misprints. The correct expansion is $\phi=\frac{Q}{r_0}\big[\frac{\Gamma(1/4)\Gamma(5/4)}{\sqrt\pi}-\frac{r}{r_0}+\frac{r^5}{10r_0^5}\big]+O(r^9)$.} 
The double copy has a horizon iff $Q>16\pi e^{3\gamma/2}\sqrt\lambda/\Gamma(\tfrac14)^4$. The black-hole star point is $\lambda=4\Qt^2/k^2$.
\subsection{Logarithmic electrodynamics}\label{sec:log}

The causal self-dual logarithmic theory~\cite{Russo:2024llm,Babaei-Aghbolagh:2025tim} has
\begin{equation}\label{lLog}
\ell=-\frac1\lambda\log\big(1-e^\gamma\lambda\tau\big),\qquad
\ld=\frac{1}{\lambda(\tm-\tau)},\qquad \tm=\frac{e^{-\gamma}}{\lambda},\qquad h_1=\tfrac12,\ h_2=\tfrac13,\ h_3=\tfrac14 .
\end{equation}
It lies in case~A with $\kappa=1$, $\mathcal A=1/\lambda$ and $n=3$. Unlike BI, $\ell(\tm)=\infty$, yet the field and the energy at the origin are finite. Gauss's law is quadratic in $E$, with causal root
\begin{equation}
E=\frac{2e^{-\gamma}Q}{r^2+\sqrt{r^4+a^4}},\qquad a^4=2\lambda e^{-\gamma}Q^2 .
\end{equation}
The potential and energy density are
\begin{equation}\label{AtLog}
A_t=\frac{r^3-r\sqrt{r^4+a^4}}{3\lambda Q}+\frac{4e^{-\gamma}Q}{3r}\,{}_2F_1\Big(\tfrac14,\tfrac12;\tfrac54;-\tfrac{a^4}{r^4}\Big),\qquad
\Ham=\frac{\sqrt{r^4+a^4}-r^2}{\lambda r^2}+\frac1\lambda\log\frac{2r^2}{r^2+\sqrt{r^4+a^4}} .
\end{equation}
Inserted in~\eqref{fexact}, these reproduce the metric of~\cite{Babaei-Aghbolagh:2025tim} term by term, including the coefficients $\tfrac{10}{9}$ and $\tfrac{16}{9}$. The near-origin data are
\begin{equation}\label{U0Log}
c_0=\frac{\Gamma(\frac14)^2}{2^{1/4}\,3\sqrt\pi}\frac{e^{-\gamma/2}\Qt^{1/2}}{\lambda^{1/4}},\qquad
c_1=-e^{-\gamma/2}\sqrt{\frac2\lambda},\qquad
\Uc_0=\frac{2^{7/4}\Gamma(\frac14)^2}{9\sqrt\pi}\frac{\Qt^{3/2}}{\lambda^{1/4}},\qquad
\Uc_1=-2\sqrt2\,\frac{\Qt}{\sqrt\lambda},
\end{equation}
and the star point is $\lambda=8\Qt^2/k^2$.

\begin{figure}[t]
\centering
\includegraphics[width=0.6\textwidth]{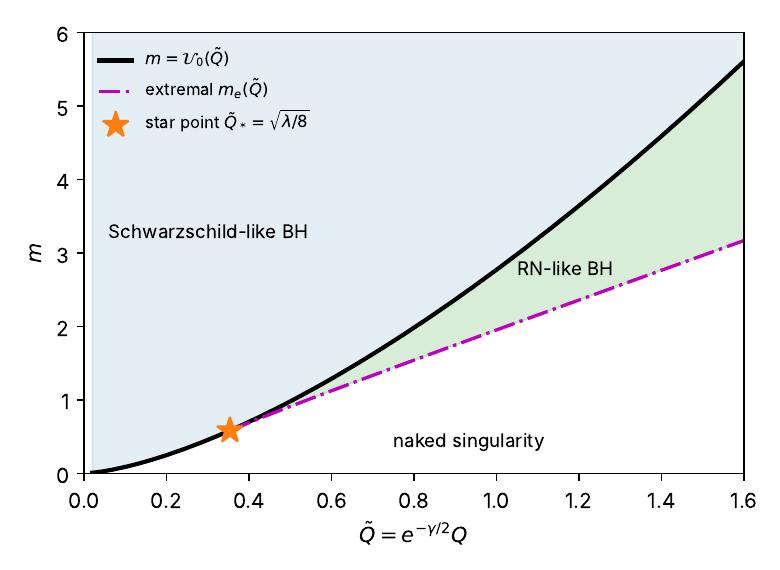}
\caption{Phase diagram of static logarithmic black holes ($k=1$, $\Lambda=0$, $\lambda=1$). The solid curve $m=\Uc_0(\Qt)$ separates the Schwarzschild-like branch. The dot-dashed curve is the extremal mass~\eqref{extremal}. The star point is at $\Qt_*=1/\sqrt8$, $m_*=0.58268$.}\label{fig:log}
\end{figure}

Several features of~\cite{Babaei-Aghbolagh:2025tim} follow directly:
\begin{itemize}
\item The two-horizon thresholds of its Figs.~1 and~3 ($k=0$, $L=1$, $r_h=1$, $\lambda=0.1$) are exactly the condition $m(r_h)=\Uc_0$, i.e.\ $2\int_0^{r_h}r^2\Ham\,\dd r=kr_h-\Lambda r_h^3/3$. This gives $\Qt=0.37729$ (Log) and $0.41407$ (GBI).
\item The extremality conditions reduce to the single equation~\eqref{extremal}; see Figure~\ref{fig:log} and Table~\ref{tab:ext}.
\item The finite self-energy means that the Cauchy horizon is not compulsory~\cite{Hale:2025ezt}. It is absent for $m>\Uc_0$ and present on the RN-like branch.
\item In the free energy (4.36) of~\cite{Babaei-Aghbolagh:2025tim}, the factors $e^{\gamma}Q^2$ should read $e^{-\gamma}Q^2$; only then does it equal $\Omega_k(2m-4\pi r_h^2T_H)$.
\end{itemize}
For the double copy, $\tilde K\simeq16e^{-\gamma}/(\lambda r^2)$, $\tilde C^2\simeq4r^2/(27\lambda^2Q^2)$, and there is a horizon iff $Q>9\sqrt2\pi e^{3\gamma/2}\sqrt\lambda/\Gamma(\tfrac14)^4$.

\begin{table}[t]
\centering\small
\begin{tabular}{@{}cccc@{}}
\toprule
$\Qt$ & $r_e$ & $m_e$ & $\Uc_0$\\
\midrule
0.36 & 0.03851 & 0.59825 & 0.59869\\
0.40 & 0.13124 & 0.69069 & 0.70119\\
0.50 & 0.28694 & 0.91084 & 0.97995\\
0.70 & 0.54160 & 1.33388 & 1.62328\\
1.00 & 0.88404 & 1.95222 & 2.77171\\
1.50 & 1.41988 & 2.96742 & 5.09195\\
\bottomrule
\end{tabular}
\caption{Extremal logarithmic black holes ($k=1$, $\Lambda=0$, $\lambda=1$). RN-like black holes exist for $m_e<m<\Uc_0$.}\label{tab:ext}
\end{table}

\subsection{\texorpdfstring{$q$}{q}-deformed electrodynamics (\texorpdfstring{$\alpha$}{alpha}-family)}

The CH-function
\begin{equation}\label{lalpha}
\ell_\alpha=\frac1\lambda\Big[1-\Big(1-\frac{e^\gamma\lambda\tau}{\alpha}\Big)^{\alpha}\Big],\qquad \tm=\frac{\alpha e^{-\gamma}}{\lambda},\qquad
h_i=\frac{(1-\alpha)(2-\alpha)\cdots(i-\alpha)}{(i+1)!\,\alpha^i},
\end{equation}
is the $q$-deformed theory of~\cite{Russo:2024xnh,Babaei-Aghbolagh:2025tim} with $q=\alpha$, and the $T=1/\lambda$ version of the family of~\cite{Russo:2026vnj}. Special cases are $\alpha=\tfrac12$ (GBI), $\alpha\to1$ (ModMax), and $\alpha\to0$ at fixed $\lambda/\alpha$ (Log). Conditions~\eqref{causal} hold for all $0<\alpha\le1$.\footnote{The range $\tfrac12\le q\le1$ quoted in~\cite{Babaei-Aghbolagh:2025tim} is therefore stronger than required by~\eqref{causal}.} This family is in case~A with $\kappa=1-\alpha$ and $n=(3-\alpha)/(1-\alpha)$.

With $y=e^\gamma\lambda\tau/\alpha\in[0,1)$, Gauss's law is solved explicitly for $r$:
\begin{equation}
\Eh=\sqrt{\frac{2\alpha y}{\lambda}},\qquad
r=\sqrt{\Qt}\Big(\frac{2\alpha y}{\lambda}\Big)^{-1/4}(1-y)^{\frac{1-\alpha}{2}},\qquad
\Ham=\frac{2\alpha y}{\lambda}(1-y)^{\alpha-1}-\frac{1-(1-y)^\alpha}{\lambda},
\end{equation}
\begin{equation}\label{Atalpha}
A_t=e^{-\gamma/2}\Big[4\sqrt{\Qt}\Big(\frac{\alpha}{8\lambda}\Big)^{1/4}y^{1/4}\,{}_2F_1\Big(\tfrac14,\tfrac{\alpha-1}{2};\tfrac54;y\Big)-r\Eh\Big].
\end{equation}
With~\eqref{fexact}, these equations give the exact black hole for every $q$ as a parametric curve. The near-origin data are
\begin{equation}\label{U0alpha}
\Uc_0=\frac43\Big(\frac{\alpha}{8\lambda}\Big)^{1/4}\frac{\Gamma(\frac14)\Gamma(\frac{3-\alpha}{2})}{\Gamma(\frac{7-2\alpha}{4})}\,\Qt^{3/2},\qquad
\Uc_1=-2\Qt\sqrt{\frac{2\alpha}{\lambda}},\qquad c_0=\frac{3e^{-\gamma/2}}{4\Qt}\Uc_0,\qquad c_1=\frac{e^{-\gamma/2}\Uc_1}{2\Qt},
\end{equation}
the star point is $\Qt_*=k\sqrt{\lambda/(8\alpha)}$, and the shifted double-copy singularity is $\tilde K\simeq16\alpha e^{-\gamma}/(\lambda r^2)$.

\subsection{Power-law and ``no maximum-\texorpdfstring{$\tau$}{tau}'' electrodynamics}

The power family
\begin{equation}\label{lbeta}
\ell_\beta=\frac1\lambda\Big[\Big(1+\frac{e^\gamma\lambda\tau}{\beta}\Big)^{\beta}-1\Big],\qquad \tm=\infty,\qquad
h_i=\frac{(\beta-1)(\beta-2)\cdots(\beta-i)}{(i+1)!\,\beta^i},
\end{equation}
with $\beta\ge1$, lies in case~B. At $\beta=\tfrac32$ it is exactly the ``no maximum-$\tau$'' theory of~\cite{Russo:2024xnh,Babaei-Aghbolagh:2025tim}, with $\lambda_{\rm NM}=2\lambda/3$. With $z=e^\gamma\lambda\tau/\beta\in[0,\infty)$:
\begin{equation}
\Eh=\sqrt{\frac{2\beta z}{\lambda}},\qquad
r=\sqrt{\Qt}\Big(\frac{2\beta z}{\lambda}\Big)^{-1/4}(1+z)^{-\frac{\beta-1}{2}},\qquad
\Ham=\frac{2\beta z}{\lambda}(1+z)^{\beta-1}-\frac{(1+z)^\beta-1}{\lambda},
\end{equation}
\begin{equation}\label{Atbeta}
A_t=e^{-\gamma/2}\Big[4\sqrt{\Qt}\Big(\frac{\beta}{8\lambda}\Big)^{1/4}z^{1/4}\,{}_2F_1\Big(\tfrac14,\tfrac{\beta-1}{2};\tfrac54;-z\Big)-r\Eh\Big].
\end{equation}
This closed form contradicts the statement in~\cite{Babaei-Aghbolagh:2025tim} that the no maximum-$\tau$ potential has no closed-form expression. Near the origin, $\Eh\simeq\hat{\mathcal C}r^{-p}$ with $\hat{\mathcal C}=[\Qt(2\beta/\lambda)^{\beta-1}]^{1/(2\beta-1)}$ and $r^2\Ham\simeq\frac{2\beta-1}{2\beta}\Qt\hat{\mathcal C}r^{-p}$. There are three regimes:
\begin{itemize}
\item $\beta>\tfrac32$: the self-energy is finite,
\begin{equation}\label{U0beta}
\Uc_0=\frac43\Big(\frac{\beta}{8\lambda}\Big)^{1/4}\frac{\Gamma(\frac14)\Gamma(\frac{2\beta-3}{4})}{\Gamma(\frac{\beta-1}{2})}\,\Qt^{3/2},
\end{equation}
but $c_1=-\infty$, $\Fc=-(m-\Uc_0)-\frac{(2\beta-1)^2}{\beta(2\beta-3)}\Qt\hat{\mathcal C}\,r^{\frac{2\beta-3}{2\beta-1}}+kr+\dots$, and there is no star point.
\item $\beta=\tfrac32$ (no maximum-$\tau$): $\Uc\simeq\tfrac43\Qt\hat{\mathcal C}\log(1/r)$ and the Cauchy horizon is compulsory.
\item $1\le\beta<\tfrac32$: $f\sim r^{-p}\to+\infty$ and the Cauchy horizon is again compulsory.
\end{itemize}

\begin{figure}[t]
\centering
\includegraphics[width=\textwidth]{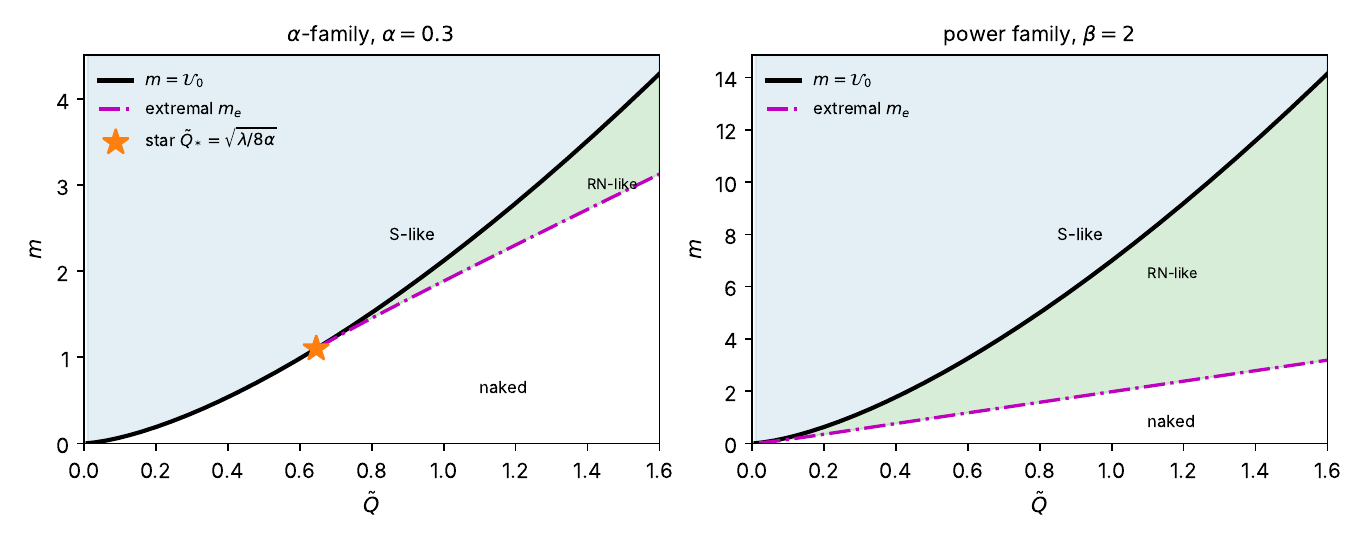}
\caption{Phase diagrams ($k=1$, $\Lambda=0$, $\lambda=1$) for the $q$-deformed theory with $\alpha=0.3$ (left) and the power family with $\beta=2$ (right). The $\alpha$-family has a star point at $\Qt_*=\sqrt{\lambda/8\alpha}=0.6455$. The power family has none, and its RN-like region extends to $\Qt\to0$.}\label{fig:fam}
\end{figure}

\subsection{Comparison}

Table~\ref{tab:crit} compares the critical points with~\eqref{crit}. Table~\ref{tab:summary} summarises the near-origin data.

\begin{table}[t]
\centering\small
\begin{tabular}{@{}lccccc@{}}
\toprule
theory & $h_1$ & $\Qt_c$ (exact) & $\Qt_c$~\eqref{crit} & $T_c$ (exact) & $T_c$~\eqref{crit}\\
\midrule
$\alpha=0.3$ & $7/6$ & 0.169010 & 0.168935 & 0.259369 & 0.259394\\
GBI ($\alpha=\tfrac12$) & $1/2$ & 0.167655 & 0.167639 & 0.259677 & 0.259682\\
Log & $1/2$ & 0.167650 & 0.167639 & 0.259679 & 0.259682\\
$\alpha=\tfrac23$ & $1/4$ & 0.167158 & 0.167153 & 0.259789 & 0.259791\\
$\alpha=\tfrac34$ & $1/6$ & 0.166994 & 0.166991 & 0.259826 & 0.259827\\
$\beta=\tfrac32$ (no max-$\tau$) & $1/6$ & 0.166990 & 0.166991 & 0.259827 & 0.259827\\
$\beta=2$ & $1/4$ & 0.167153 & 0.167153 & 0.259790 & 0.259791\\
$\beta=3$ & $1/3$ & 0.167316 & 0.167315 & 0.259754 & 0.259755\\
\bottomrule
\end{tabular}
\caption{Critical points for $k=1$, $L=1$, $\lambda=0.02$, compared with the universal first-order formula~\eqref{crit}.}\label{tab:crit}
\end{table}

\begin{table}[t]
\centering\footnotesize\setlength{\tabcolsep}{3pt}
\renewcommand{\arraystretch}{1.5}
\resizebox{\textwidth}{!}{%
\begin{tabular}{@{}lcccc@{}}
\toprule
& $q$-deformed ($\alpha$) & GBI ($\alpha=\frac12$) & Log ($\alpha\to0$) & power ($\beta$)\\
\midrule
$\tm$ & $\alpha e^{-\gamma}/\lambda$ & $e^{-\gamma}/2\lambda$ & $e^{-\gamma}/\lambda$ & $\infty$\\
case & A, $n=\frac{3-\alpha}{1-\alpha}$ & A, $n=5$ & A, $n=3$ & B, $p=\frac{2}{2\beta-1}$\\
$h_1$ & $\frac{1-\alpha}{2\alpha}$ & $\frac12$ & $\frac12$ & $\frac{\beta-1}{2\beta}$\\
$\Uc_0\lambda^{1/4}/\Qt^{3/2}$ & $\frac43(\frac{\alpha}{8})^{\frac14}\frac{\Gamma(\frac14)\Gamma(\frac{3-\alpha}{2})}{\Gamma(\frac{7-2\alpha}{4})}$ & $\frac{\Gamma(\frac14)^2}{3\sqrt\pi}$ & $\frac{2^{7/4}\Gamma(\frac14)^2}{9\sqrt\pi}$ & $\frac43(\frac{\beta}{8})^{\frac14}\frac{\Gamma(\frac14)\Gamma(\frac{2\beta-3}{4})}{\Gamma(\frac{\beta-1}{2})}$, $\beta>\frac32$\\
$\Uc_1/\Qt$ & $-2\sqrt{2\alpha/\lambda}$ & $-2/\sqrt\lambda$ & $-2\sqrt{2/\lambda}$ & $-\infty$\\
star point ($k=1$) & $\lambda=8\alpha\Qt^2$ & $\lambda=4\Qt^2$ & $\lambda=8\Qt^2$ & none\\
Cauchy horizon & not compulsory & not compulsory & not compulsory & compulsory iff $\beta\le\frac32$\\
residual DC singularity $\tilde K$ & $16\alpha e^{-\gamma}/\lambda r^2$ & $8e^{-\gamma}/\lambda r^2$ & $16e^{-\gamma}/\lambda r^2$ & $\propto r^{-2-2p}$\\
\bottomrule
\end{tabular}}
\caption{Near-origin data of the double copy and of the Einstein--NED black hole for the families studied.}\label{tab:summary}
\end{table}

\section{Discussion}\label{sec:disc}

We have shown that the short-distance physics of point charges in causal self-dual NED is controlled by very little data from the CH-function: the functional $\Ic[\ell]$ and the end point $\tm$. These enter as the coefficients $c_0$ and $c_1$ of the electrostatic potential. In the Kerr--Schild double copy they fix the curvature singularity and the horizon. In Einstein gravity they fix the self-energy $\Uc_0=\frac43Qc_0$ and the next coefficient $\Uc_1=2Qc_1$. Through these they fix the interior branch structure, the star point and the naked-singularity region. Lemma~\ref{lem:virial} makes this bridge exact, and it also renders the Einstein--NED equations algebraic in $(A_t,\Ham)$ for any NED.

We stress that this bridge relates \emph{data}, not solutions. The Kerr--Schild double copy does \emph{not} generate solutions of the Einstein--NED system. The double-copy metric has $f_{\rm DC}=1-A_t(r)$, whereas the Einstein--NED black hole has the different function~\eqref{fexact}. The two already differ for Maxwell theory: the double copy of the Coulomb charge is the Schwarzschild metric with $M_{\rm DC}=Q/2$, in which the charge has been converted into mass and no electromagnetic field survives, while Einstein--Maxwell gives Reissner--Nordstr\"om, with independent $m$ and $Q$. For non-conformal theories, the effective source of the double copy (Proposition~\ref{prop:Ricci}) is an anisotropic fluid with $\rho_{\rm eff}\propto(\phi-rE)/r^2$. This is not the NED energy density $\Ham=ED-\Lg$; for Maxwell, $\phi-rE=0$ while $\Ham=Q^2/2r^4$. Conversely, writing the Einstein--NED black hole in Kerr--Schild form, with $\phi_{\rm BH}=1-f$, gives a single copy that solves the \emph{Maxwell} equations with an extended charge density, not the NED equations, as for Reissner--Nordstr\"om in~\cite{Monteiro:2014cda}. The natural gravitational counterpart of the double copy is not Einstein--NED but the graviton--dilaton--$B$-field theory, which is why the general double copy of a point charge is the JNW solution~\cite{Kim:2019jwm}. What the two constructions share is that the same two numbers, $c_0=\sqrt Q\,\Ic[\ell]$ and $c_1=-\sqrt{2\tm}$, govern their behaviour near the origin: the double-copy singularity and horizon on one side, and $\Uc_0=\tfrac43Qc_0$, $\Uc_1=2Qc_1$ on the other.

Three lessons stand out.

\emph{First}, the Ricci tensor of the double copy is a direct probe of conformal symmetry breaking, $R^t{}_t\propto\ldd\propto-\dd\Theta/\dd\tau$. Along the root-$T\bar T$ flow from Maxwell to ModMax the double copy stays Schwarzschild, while the irrelevant $T\bar T$ deformation sources an effective anisotropic fluid of positive energy.

\emph{Second}, the no-go theorem~\ref{thm:nogo} sharpens the conclusion of~\cite{Pasarin:2020qoa}. It is not a peculiarity of Born--Infeld that its double copy is singular. Every causal NED fails, and BI-type theories already give the mildest possible singularity, with a regular Weyl tensor in the shifted gauge. Regular double copies therefore require leaving the minimal Kerr--Schild map. Natural options are the dilatonic (JNW) double copy~\cite{Kim:2019jwm}, derivative corrections beyond BI~\cite{Tseytlin:1995uq,Pasarin:2020qoa}, or non-Kerr--Schild constructions. In the amplitudes language~\cite{Bern:2019prr,Adamo:2022dcm}, Born--Infeld is itself a double copy rather than a single copy, which is a further reason to expect a modified map.

\emph{Third}, the Einstein--NED results organise the phenomenology of~\cite{Babaei-Aghbolagh:2025tim,Hale:2025ezt,Russo:2026vnj} into a simple dictionary. Finite $\tm$ (case~A) implies a finite self-energy, an optional Cauchy horizon and a star point. Case~B splits at $\beta=\tfrac32$, which is precisely the ``no maximum-$\tau$'' theory: above it the self-energy is finite but there is no star point; at and below it the Cauchy horizon is compulsory. The van der Waals critical point of AdS black holes is, by contrast, a weak-field phenomenon. At $O(\lambda)$ it depends on $h_1$ alone, so theories as different as $q=\tfrac23$ and $\beta=2$ are thermodynamically indistinguishable near criticality but have different interiors.

Several directions are open:
\begin{itemize}
\item Extend the analysis to dyonic and magnetic configurations, where $U\ne0$ and the full CH structure enters.
\item Extend it to rotating Kerr--Schild solutions.
\item Study the flow of $\Ic[\ell_\lambda]$ and $\tm(\lambda)$ under exact $T\bar T$-type deformations~\cite{Pezhman:2026dsr}, which governs how the black-hole/naked-singularity boundary moves along the flow.
\item Push the critical-point analysis to $O(\lambda^2)$, where the strong-field differences first appear.
\item Find holographic interpretations of $c_0$ and $c_1$.
\item Clarify the relation to regular black holes sourced by NED~\cite{Ayon-Beato:1998hmi,Balart:2014cga}, which evade our no-go statements because they violate one of its assumptions (causality, duality invariance, or a point-like source). The finite-self-energy problem in BI-type theories~\cite{Dehghani:2021fwb} suggests a field-theoretic counterpart of these questions.
\end{itemize}

\begin{acknowledgments}
The author is grateful to Komeil Babaei Velni, Song He, Karapet Mkrtchyan, Alireza Pahlavan, Jorge G. Russo, Dmitri Sorokin, and Roberto Tateo for their interest in this work and the fruitful discussions that followed. This work was conducted as part of the PostDoc Program on {\it Exploring TT-bar Deformations: Quantum Field Theory and Applications}, sponsored by Ningbo University. 
\end{acknowledgments}

\appendix
\section{Special functions and \texorpdfstring{$\Gamma(\frac14)$}{Gamma(1/4)}}\label{app:special}

We use
\begin{equation}
B(a,b)=\int_0^1t^{a-1}(1-t)^{b-1}\dd t=\int_0^\infty\frac{x^{a-1}\dd x}{(1+x)^{a+b}}=\frac{\Gamma(a)\Gamma(b)}{\Gamma(a+b)},
\end{equation}
\begin{equation}
B(y;a,b)=\int_0^yt^{a-1}(1-t)^{b-1}\dd t=\frac{y^a}{a}\,{}_2F_1(a,1-b;a+1;y).
\end{equation}
All Beta functions in this paper have $a=\tfrac14$, because $\Ic[\ell]$ has the integrand $\tau^{-3/4}$. For the two families,
\begin{equation}
B\Big(\tfrac14,\tfrac{3-\alpha}{2}\Big)=\Gamma(\tfrac14)\frac{\Gamma(\frac{3-\alpha}{2})}{\Gamma(\frac{7-2\alpha}{4})},\qquad
B\Big(\tfrac14,\tfrac{2\beta-3}{4}\Big)=\Gamma(\tfrac14)\frac{\Gamma(\frac{2\beta-3}{4})}{\Gamma(\frac{\beta-1}{2})},
\end{equation}
and the integrals $\Jc$ in~\eqref{Atalpha} and~\eqref{Atbeta} are
\begin{align}
\int_0^yx^{-3/4}(1-x)^{\frac{1-\alpha}{2}}\dd x&=B\big(y;\tfrac14,\tfrac{3-\alpha}{2}\big)=4y^{1/4}\,{}_2F_1\big(\tfrac14,\tfrac{\alpha-1}{2};\tfrac54;y\big),\\
\int_0^zx^{-3/4}(1+x)^{-\frac{\beta-1}{2}}\dd x&=4z^{1/4}\,{}_2F_1\big(\tfrac14,\tfrac{\beta-1}{2};\tfrac54;-z\big)\ \xrightarrow{z\to\infty}\ B\big(\tfrac14,\tfrac{2\beta-3}{4}\big).
\end{align}
Reflection, duplication and triplication,
\begin{equation}
\Gamma(z)\Gamma(1-z)=\frac{\pi}{\sin\pi z},\qquad
\Gamma(z)\Gamma(z+\tfrac12)=2^{1-2z}\sqrt\pi\,\Gamma(2z),
\end{equation}
\begin{equation}
\Gamma(z)\Gamma(z+\tfrac13)\Gamma(z+\tfrac23)=2\pi\,3^{\frac12-3z}\Gamma(3z),
\end{equation}
give in particular
\begin{equation}
\Gamma(\tfrac34)=\frac{\pi\sqrt2}{\Gamma(\frac14)},\qquad
\frac{\Gamma(\frac18)}{\Gamma(\frac38)}=\frac{2^{3/4}\sin\frac{3\pi}8}{\sqrt\pi}\Gamma(\tfrac14),
\end{equation}
\begin{equation}
\Gamma(\tfrac16)=\frac{\sqrt3\,\Gamma(\frac13)^2}{2^{1/3}\sqrt\pi},\qquad
\Gamma(\tfrac5{12})=\frac{\sqrt{\pi(\sqrt6-\sqrt2)}}{3^{1/8}}\frac{\Gamma(\frac14)}{\Gamma(\frac13)} .
\end{equation}
For the power family, everything reduces to $\Gamma(\tfrac14)$ and $\pi$ whenever $2\beta\in\mathbb Z$. Table~\ref{tab:gamma} lists the special values, with $\Uc_0=u\,\Qt^{3/2}\lambda^{-1/4}$ and $c_0=\tfrac34e^{-\gamma/2}u\,\Qt^{1/2}\lambda^{-1/4}$.

\begin{table}[h]
\centering\footnotesize
\renewcommand{\arraystretch}{1.6}
\resizebox{\textwidth}{!}{%
\begin{tabular}{@{}llll@{}}
\toprule
theory & Beta function & in terms of $\Gamma$ & $u$\\
\midrule
GBI & $B(\frac14,\frac54)$ & $\frac{\Gamma(\frac14)^2}{2\sqrt\pi}$ & $\frac{\Gamma(\frac14)^2}{3\sqrt\pi}\approx2.4721$\\
$q=\frac34$ & $B(\frac14,\frac98)$ & $\frac{\sqrt{2+\sqrt2}}{2^{1/4}}\frac{\Gamma(\frac14)^2}{3\sqrt\pi}$ & $\frac{\sqrt{4+2\sqrt2}}{3^{7/4}}\frac{\Gamma(\frac14)^2}{\sqrt\pi}\approx2.8339$\\
$q=\frac23$ & $B(\frac14,\frac76)$ & $\frac{2^{2/3}3^{5/8}}{5}\frac{\Gamma(\frac13)^3}{\pi\sqrt{\sqrt6-\sqrt2}}$ & $\frac{2^{13/6}}{5\cdot3^{5/8}}\frac{\Gamma(\frac13)^3}{\pi\sqrt{\sqrt6-\sqrt2}}\approx2.7181$\\
Log & $B(\frac14,\frac32)$ & $\frac{\sqrt2\,\Gamma(\frac14)^2}{3\sqrt\pi}$ & $\frac{2^{7/4}\Gamma(\frac14)^2}{9\sqrt\pi}\approx2.7717$\\
$\beta=2$ & $B(\frac14,\frac14)$ & $\frac{\Gamma(\frac14)^2}{\sqrt\pi}$ & $\frac{2\sqrt2}{3}\frac{\Gamma(\frac14)^2}{\sqrt\pi}\approx6.9922$\\
$\beta=\frac52$ & $B(\frac14,\frac12)$ & $\frac{\Gamma(\frac14)^2}{\sqrt{2\pi}}$ & $\frac{\sqrt2\,5^{1/4}}{3}\frac{\Gamma(\frac14)^2}{\sqrt\pi}\approx5.2279$\\
$\beta=3$ & $B(\frac14,\frac34)$ & $\pi\sqrt2$ & $2^{7/4}3^{-3/4}\pi\approx4.6357$\\
$\beta=4$ & $B(\frac14,\frac54)$ & $\frac{\Gamma(\frac14)^2}{2\sqrt\pi}$ & $\frac{2^{3/4}}{3}\frac{\Gamma(\frac14)^2}{\sqrt\pi}\approx4.1576$\\
\bottomrule
\end{tabular}}
\caption{Self-energy coefficients in terms of $\Gamma$-functions.}\label{tab:gamma}
\end{table}


\bibliographystyle{JHEP}
\bibliography{main1_references}

\end{document}